\documentclass[a4paper,11pt]{article}
\usepackage{pos}

\usepackage[utf8]{inputenc}	
\usepackage{siunitx}
\usepackage{xspace}
\usepackage{amsmath,bm,mathrsfs,mathtools,dsfont}
\usepackage{slashed} 
\usepackage{xfrac}

\usepackage[shortlabels]{enumitem}
\setlist{itemsep=.1em,topsep=.5em}
\SetEnumerateShortLabel{i}{\textit{\roman*}}

\usepackage{tabularx}
\newcolumntype{Y}{>{\centering\arraybackslash}X}

\renewcommand{\L}{\mathcal{L}}
\newcommand{\LL}{\mathrm{L}}
\newcommand{\RR}{\mathrm{R}}
\newcommand{\U}{\mathrm{U}}
\newcommand{\SU}{\mathrm{SU}}
\newcommand{\rep}[1]{\bm{#1}}
\newcommand{\repbar}[1]{\overline{\bm{#1}}}
\newcommand{\eminus}{\vcenter{\hbox{\scalebox{0.6}[1]{$ - $}}}}

\newcommand{\gmu}{$ (g-2)_\mu $\xspace}
\newcommand{\RK}{$ R_{K^{(\ast)}} $\xspace}

\title{Addressing the Muon Anomalies With Muon-Flavored Leptoquarks}
\ShortTitle{Addressing muon anomalies with muoquarks}

\author{Admir Greljo}
\author{Peter Stangl}
\author*{Anders Eller Thomsen}

\affiliation{Albert Einstein Center for Fundamental Physics, Institute for Theoretical Physics,\\
  University of Bern, CH-3012 Bern, Switzerland}

\emailAdd{admir.greljo@unibe.ch}
\emailAdd{peter.stangl@unibe.ch}
\emailAdd{anders.thomsen@unibe.ch}

\abstract{Significant deviations from Standard Model (SM) predictions have been observed in $ b \to s \mu^+ \mu^-$ decays and in the muon $g-2$. Scalar leptoquark extensions of the SM are known to be able to address these anomalies, but generically give rise to lepton flavor violation (LFV) or even proton decay. We propose new muon flavored gauge symmetries as a guiding principle for leptoquark models that preserve the global symmetries of the SM and explain the non-observation of LFV. A minimal model is shown to easily accommodate the anomalies without encountering other experimental constraints. This talk is mainly based on Ref.~\cite{Greljo:2021xmg}.}

\FullConference{%
  *** The European Physical Society Conference on High Energy Physics (EPS-HEP2021), ***\\
  *** 26-30 July 2021 ***\\
  *** Online conference, jointly organized by Universität Hamburg and the research center DESY ***
}

\begin{document}
\maketitle

\section{Introduction}
The discovery of the muon in 1936 sparked the beginning of modern particle physics. It is, thus, almost poetic that experimental evidence is hinting at new physics coupled precisely to the muon. While nothing is conclusive as of yet, it would indeed be fitting if once again the muon were to spark a new era in physics, this time the era of physics Beyond the Standard Model (BSM). 

A whole class of rare $ B $ meson decay measurements, through $ b \to s \mu^+ \mu^- $ transitions, seem to deviate from SM theory predictions. Most prominent among these results are the theoretically clean~\cite{Hiller:2003js,Bordone:2016gaq} lepton flavor universality ratios, $ R_{K^{(\ast)}} $, indicating a deficit of muons compared to electrons~\cite{Aaij:2014ora,Aaij:2017vbb}. Notably, the most recent $ R_{K} $ measurement deviates from the SM with $ 3.1\sigma $~\cite{Aaij:2021vac}. Also the world average of $ \mathrm{BR}(B_s \to \mu^+ \mu^-) $~\cite{Altmannshofer:2021qrr} and various less theoretically clean angular observables deviate from the SM~\cite{Aaij:2015oid}. A coherent picture emerges from all of these observables showing a strong preference for new physics, in the form of an effective left-handed semi-leptonic current, over the SM~\cite{Isidori:2021vtc}.      

The measurement of the anomalous magnetic moment of the muon has long been in tension with the SM theory prediction. The first result from the Fermilab Muon g-2 experiment~\cite{Abi:2021gix} confirms the old Brookhaven result~\cite{Bennett:2006fi} and yields a combined tension of $ 4.2\sigma $ with the improved SM prediction from the Muon g-2 Theory Initiative~\cite{Aoyama:2020ynm} (a recent lattice study suggests a significantly smaller tension~\cite{Borsanyi:2020mff} but has yet to be verified by other groups). 
Proceeding under the assumption that the deviations presented here are due to genuine new physics (NP) beyond the SM, the $ (g-2)_\mu $ tension suggests an emphasis on muons, and we consider \gmu and $ b \to s \mu^+ \mu^- $ a combined set of muon anomalies. 

Much has been made of the muon anomalies, but just as important, albeit less glamorous, are all the processes we do not observe. The vanilla SM has an accidental $ \U(1)_B \times \U(1)_{L_e} \times \U(1)_{L_\mu} \times \U(1)_{L_\tau} $ symmetry, ensuring conservation of baryon number and individual lepton numbers for all three generations. The non-observation of proton decay is famously enough to push the GUT scale to $ \SI{e15}{GeV} $ in grand unified theories. The situation is slightly less severe for lepton number conservation. Nevertheless, the effective dipole operator
	\begin{equation}
	\mathcal{L} \supset - \frac{ e \,v}{(4 \pi)^2 \Lambda_{ij}^2}\, \bar \ell^i_{\LL} \sigma^{\mu \nu} \ell^j_{\RR } F_{\mu \nu} + \; {\rm H.c.}
	\end{equation}
is severely constrained, in particular from the non-observation of $ \mu \to e \gamma $, ensuring $ \Lambda_{e\mu} > \SI{3e4}{TeV} $~\cite{Calibbi:2021qto}.\footnote{Considering the SM as an effective theory, the observation of neutrino oscillations requires \emph{some} small lepton-number and flavor violation easily consistent with these bounds.} This is to be compared to the best fit value $ \Lambda_{\mu\mu} = \SI{14}{TeV}$ for \gmu. It is quite clear from this that a heavy NP explanation cannot be flavor anarchic. Rather, it could have hierarchical couplings to the lepton generations (although the $ \tau $ flavor violation is constraining, $ \Lambda_{\mu\tau} > \SI{30}{TeV} $) or, as we explore here, muon-specific NP. 

An undeniable solution to these constraints is to equip the BSM model with a symmetry that ensures that the resulting low-energy effective theory possesses the same accidental symmetries as the SM.  
Our philosophy here is to construct a simple UV model, even an entire class of models, to explain the muon anomalies while preserving the accidental symmetries of the SM.

\section{Introducing the Muoquark}
Leptoquarks (LQ) have long been considered as explanations for flavor anomalies, e.g. in Refs.~\cite{Gripaios:2009dq,Hiller:2014yaa,Bauer:2015knc,Barbieri:2015yvd,DiLuzio:2017vat,Angelescu:2018tyl}. They can mediate the semileptonic transitions at tree-level, whereas the heavily constrained $ \Delta F=2 $ transitions are loop suppressed. Loops with LQs and top quarks can also facilitate lepton dipoles with chiral enhancement, suggesting a common NP scale for both $ b \to s \mu^+ \mu^- $ and \gmu anomalies of order $ \SI{10}{TeV} $.

\subsection{Scalar leptoquarks}
No single scalar LQ representation can account for the $ b \to s \mu^+ \mu^- $ and \gmu anomalies simultaneously. $ S_3 \sim (\repbar{3},\, \rep{3},\, 1/3) $ is unique among the scalar LQs in that it can mediate the left-handed semileptonic current needed for $ b \to s \mu^+ \mu^- $ at tree level. Either $ S_1 \sim (\repbar{3},\, \rep{1},\, 1/3) $ or $ R_2 \sim (\rep{3},\, \rep{2},\, 7/6) $ are separately able to account for \gmu~\cite{Crivellin:2017zlb,Gherardi:2020qhc}. $S_1/R_2$ and $S_3$ give sub-leading contributions to $R_K$ and $(g-2)_\mu$, respectively. 

One of the main issues with scalar LQs is that they tend to violate the global symmetries of the SM. Schematically, the Yukawa couplings
	\begin{equation}
	\L \supset -y (QLS)  - z(QQ S^\dagger)  
	\end{equation}
are generically possible for the scalar LQs under consideration. Here $ Q $ and $ L $ denote generic quarks and leptons, respectively. One will quickly find that it is impossible to assign a baryon number to $ S $, that is, integrating out $ S $ will give rise to a $ B $-violating 4-fermion operator mediating proton decay~\cite{Arnold:2013cva,Assad:2017iib}. In fact, even a suppression by the Planck scale, $ z \sim v/ M_\mathrm{Pl} $, is typically insufficient. Invoking global symmetries in the LQ theory to exclude $ z $ might well be insufficient given that gravity is commonly held to violate global symmetries~\cite{Banks:2010zn}.   

Compared to $ B $ violation, LFV processes are less constraining, nevertheless, they limit how strongly any one LQ can couple to multiple generations of leptons.  
The idea is now to introduce a new gauge symmetry along with the LQs to ensure that the low-energy theory reproduces the global symmetries of the SM.

\subsection{$ \U(1) $ symmetry}
As a minimal symmetry, we impose a $ \U(1)_X $ gauge symmetry to rescue the scalar LQ models~\cite{Hambye:2017qix,Davighi:2020qqa}. Seeing as we do not observe any new massless gauge-boson coupling to ordinary matter fields, we will assume a new SM-singlet scalar field $ \Phi $ to break the $ \U(1)_X $ symmetry.

We require that the $ \U(1)_X $ symmetry satisfy the conditions:
	\begin{enumerate}[i)]
	\item the leptonic $ X $ charges should be generation-dependent allowing for the coupling 
	$ Q L_i S $ for $ i=\mu $ but not for $ i=e,\tau $;
	\item the charges of $ S $ and $ Q $ must forbid all $ QQ S^\dagger $ operators;
	\item the charge of $ \Phi $ should forbid the dimension-5 operators $ QQ S^\dagger \Phi^{(\ast)} $;
	\item a remnant (approximate) individual LF symmetry.
	\end{enumerate}	
These conditions ensure the restoration of the accidental symmetries of the SM and have been explored in full detail in Ref.~\cite{Greljo:2021npi}. Furthermore the LQs couple exclusively to muons, and we may properly refer to them as \emph{muoquarks}. 
There are additional properties we might wish for the $ \U(1)_X $ symmetry to posses to ensure a simple UV model, namely,
	\begin{enumerate} [i)]
	\setcounter{enumi}{4}
	\item universal quark charges (proportional to $ B $) to allow directly for the full SM quark Yukawa matrices and avoid flavor-violating $ X_\mu $ currents;
	\item a $ \Phi $ charge allowing for Yukawa couplings $ \bar{\nu}^c_\RR \nu_\RR  \Phi^{(\ast)} $ to populate otherwise forbidden Majorana mass entries and produce the full PMNS matrix with no additional fields.  
	\end{enumerate}
The $ \U(1)_{B-3L_\mu} $ symmetry emerges as a candidate satisfying all of the conditions listed here.

\subsection{Model}
We proceed to introduce a simple muoquark models with a gauged $ \U(1)_{B-3L_\mu} $ symmetry. The model consists of the SM extended with right-handed neutrinos, $ S_1 $ and $ S_3 $ LQs, an SM singlet scalar $ \Phi $, and the gauge field $ X_\mu $ of the new symmetry. The charges of the fields are shown in Table~\ref{tab:model}, while the full Lagrangian is given by 
	\begin{equation} \label{eq:Lagrangian}
	\begin{split}
	\L =&\, \L_{\mathrm{SM} -V_H} + \bar \nu^{i}_\RR i \slashed{D} \nu_\RR^i + |D_\mu \Phi|^2 + |D_\mu S_1|^2 + |D_\mu S_3|^2 - \tfrac{1}{4} X_{\mu \nu}^2 +\tfrac{1}{2}\varepsilon_{BX} B_{\mu \nu} X^{\mu \nu} \\
	&- \big(\eta^{3\LL}_i \, \overline{q}^{c\,i}_\LL  \ell^2_\LL \, S_3 + \eta^{1\LL}_i \overline{q}^{c\,i}_\LL \ell^2_\LL S_1 + \eta^{1\RR}_i \overline{u}^{c\,i}_\RR \mu_\RR S_1 + \tilde{\eta}^{1\RR}_i \overline{d}^{c\,i}_\RR \nu_{\mu,\RR} S_1  + \mathrm{H.c.} \big)
	 \\
	&- \big( y_\nu^{i j} \bar \ell^i_\LL \tilde H \nu_\RR^j + M^{ij}_\RR \bar \nu^{c i}_\RR \nu_\RR^j + y_\Phi^{ij} \Phi\,  \bar \nu^{c i}_\RR \nu_\RR^j + \mathrm{H.c.} \big) - V(H, \Phi, S_1, S_3),
	\end{split}
	\end{equation}
for a suitable scalar potential $ V $. 

We take the scalar potential to be such that $ \Phi $ develops a VEV, $ v_\Phi = \langle \Phi \rangle $. As long as we arrange for $ V $ to reproduce the standard Higgs potential to recover ordinary electroweak symmetry breaking, the remaining part of the potential is of limited phenomenological relevance. Here we will pursue the decoupling limit $ g_X \to 0 $ and $m_X \to \infty$, so as not to consider the $ X_\mu $ phenomenology explicitly. 

\begin{table}
	\centering \small 
	\begin{tabularx}{.65\textwidth}{| Y | Y Y Y Y|} 
	\hline \hline 
	Fields & $ \SU(3)_c $ & $ \SU(2)_\LL $ & $ \U(1)_Y $ & $ \U(1)_{B-3L_\mu} $ \\
	\hline 
	$ q_\LL $ & $ \rep{3} $ & $ \rep{2} $ & $ \sfrac{1}{6}$ & $ \sfrac{1}{3} $\\
	$ u_\RR $ & $ \rep{3} $ & & $ \sfrac{2}{3}$ & $ \sfrac{1}{3} $\\
	$ d_\RR $ & $ \rep{3} $ & & $ \eminus \sfrac{1}{3}$ & $ \sfrac{1}{3} $\\
	$ \ell_\LL $ & & $ \rep{2} $ & $ \eminus \sfrac{1}{2}$ & $ \{0,\, \eminus3,\, 0 \} $\\
	$ e_\RR $ & & & $ \eminus 1$ & $ \{0,\, \eminus3,\, 0 \} $\\
	$ \nu_\RR $ & &  & $ 0$ & $ \{0,\, \eminus3,\, 0 \} $\\ \hline 
	$ H  $ & & $ \rep{2} $ & $ \sfrac{1}{2} $ & $ 0 $\\
	$ S_3  $ & $ \repbar{3} $ & $ \rep{3} $ & $ \sfrac{1}{3} $ & $ \sfrac{8}{3} $ \\
	$ S_1  $ & $ \repbar{3} $ &  & $ \sfrac{1}{3} $ & $ \sfrac{8}{3} $ \\
	$ \Phi $ & & & $ 0 $ & $ 3 $ \\
	\hline \hline 
	\end{tabularx}
	\caption{Field content of the $ \U(1)_{B-3L_\mu} $ muoquark model with their representation under the gauge groups.}
	\label{tab:model}
\end{table}

After $ \U(1)_{B-3L_\mu} $ symmetry breaking, there are remnant $ \U(1)_{L_\mu} $ and $ \U(1)_B $ global symmetries, the latter of which is exact (but anomalous).\footnote{All terms in the potential involve $ \Phi $ only through the combination $ | \Phi|^2 $.} The $ \U(1)_{L_\mu} $ symmetry is explicitly broken by the right-handed neutrino Majorana mass term due to the $ y_\Phi^{ij} v_\Phi $ term; however, with heavy right-handed neutrino masses at the multi-TeV scale, this is mediated only to the effective Weinberg operator at low scales.

\subsection{Fit}

\begin{figure}
	\centering
	\includegraphics[width=.5\textwidth]{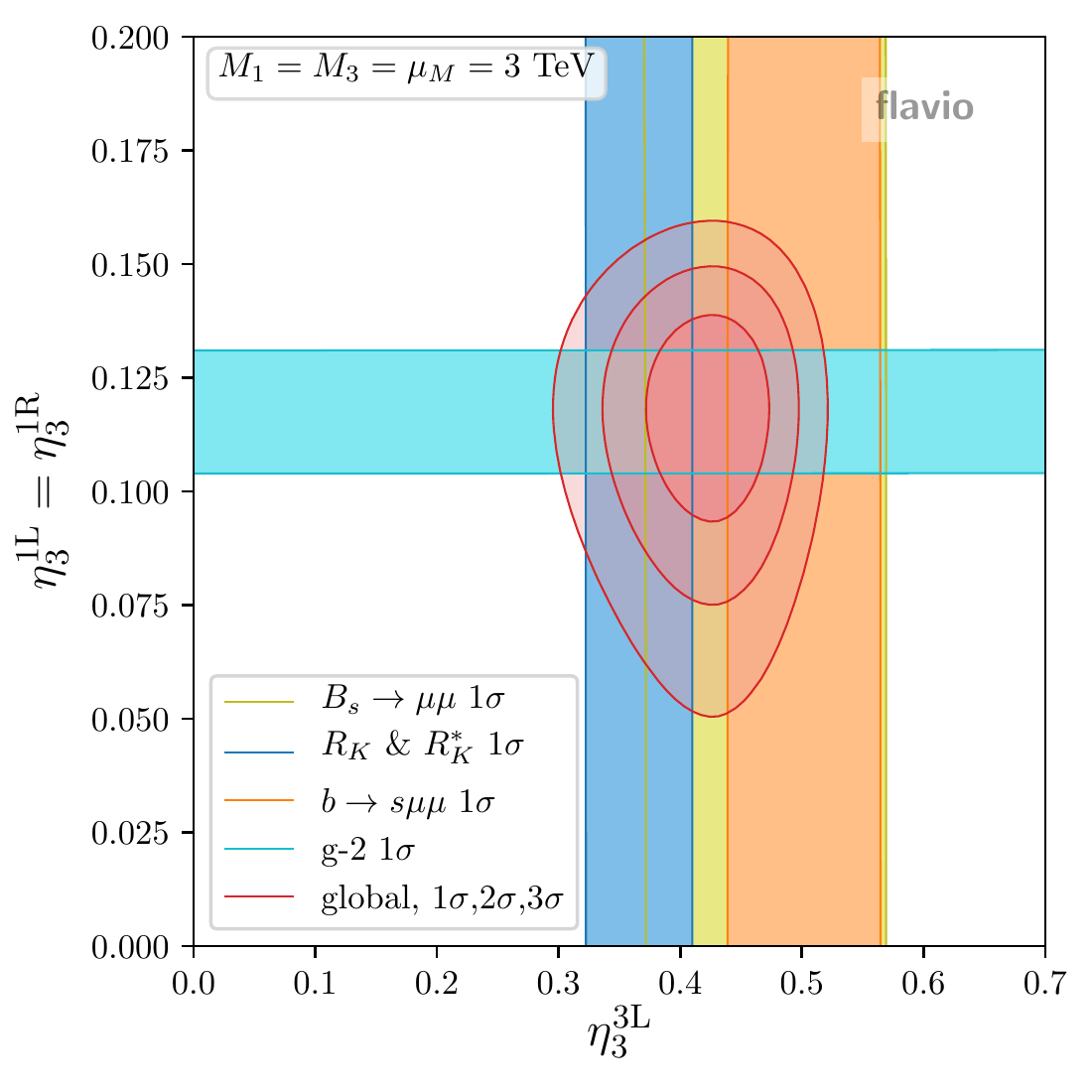}
	\caption{The preferred muoquark Yukawa couplings from the global fit to low-energy data, with the colored bands corresponding to best fit regions of individual observables. Here we choose $ \eta^{3\LL}_i = (V_{td}, \, V_{ts},\, 1) \, \eta^{3\LL}_{3} $, $ \eta^{1\LL}_i = (V_{td}, \, V_{ts},\, 1)\, \eta^{1\LL}_{3} $, and $  \eta^{1\RR}_i = (0,\, 0,\, 1)\, \eta^{1\RR}_3 $.
The muoquark masses are set to $ M_1 = M_3 = \SI{3}{TeV} $. }
	\label{fig:muon_fit}
\end{figure}

Whereas the lepton couplings to the muoquarks are completely fixed by the $ \U(1)_{B-3L_\mu} $ symmetry, there is no a priori structure to the quark couplings. Nevertheless, a reasonable structure can be assigned from a supposition that the LQ couplings are generated by the same UV physics that is responsible for the structure of the SM Yukawa couplings. That is, we proceed with the assumption that the approximate $ \U(2)^3 $ symmetry of the SM quark sector~\cite{Barbieri:2011ci,Kagan:2009bn} is broken by the minimal set of spurions also in the LQ couplings.    

To verify that our proposed model provides a good explanation of the muon anomalies while not running afoul of any other EW or flavor physics constraints, we perform a global fit in a subspace of the Yukawa couplings. To this end, we employ the full 1-loop matching of $ S_1 + S_3 $ LQs~\cite{Gherardi:2020det} to obtain the low-energy effective theory of the model (in the limit of decoupled $ \Phi $ and $ X_\mu $). We implement the result in a code that interfaces with the \texttt{smelli} package~\cite{Aebischer:2018iyb,Stangl:2020lbh}. The SM effective theory is run down to the EW scale, matched to the low-energy EFT, and run to the QCD scale thanks to the \texttt{wilson} package~\cite{Aebischer:2018bkb}; and \texttt{flavio}~\cite{Straub:2018kue} is used to compute a large number of electroweak and flavor observables (see Ref.~\cite{smelli} for a complete list).   

As a benchmark we take both muoquarks to have mass $ M_{1,3} = \SI{3}{TeV} $, well beyond current direct searches~\cite{Aad:2020iuy,ATLAS:2020qoc}. This constitutes a rather pessimistic scenario (from the perspective of NP searches) that is even beyond projections for HL-LHC searches in the high-$ p_T $ lepton tails~\cite{Greljo:2017vvb}. One might be lucky that nature realizes this scenario with low LQ masses, but our point here is to emphasize that NP explanations of the muon anomalies reproducing the SM global symmetries can come with no further signs of NP at LHC energies. We plot the best-fit regions to the muoquark Yukawa couplings in Fig.~\ref{fig:muon_fit}, showing a best fit favored with $ \Delta \chi^2 \simeq 62 $ over the SM. The vertical bands show the $ 1\sigma $ range for \RK, $ \mathrm{BR}(B_s\to \mu^-\mu^+) $, and the less theoretically clean $ b\to s \mu^- \mu^+ $ branching ratios and angular observables, while the horizontal band is the preferred region for \gmu. The plot indicates a clear factorization of the role of the muoquarks in explaining the anomalies. We find no strong constraints from any other flavor observables even when varying order-1 coefficients in front of the spurion couplings of the muoquarks to lighter generation quarks.

\subsection{Theory constraints}
The muoquark model presented here is renormalizable, and the RG flow is free from Landau poles up to the Planck scale. The potential can also be made stable all the way up to the Planck scale, so there is no sign of the theory breaking down before then, giving no further signs of NP. 

Theory constraints do, however, play an important role in setting a soft upper limit on the muoquark scale. The NP contribution to the muon dipole has a chiral enhancement due to the top quark entering the loop. The same chiral enhancement can be found in the $ S_1 $ matching correction to the muon Yukawa~\cite{Gherardi:2020det,Capdevilla:2020qel,Capdevilla:2021rwo}: 
	\begin{equation}
	\delta y_\mu = -\dfrac{3}{(4\pi)^2} \left(1 + \ln \dfrac{\mu_M^2}{M_1^2} \right) \eta_i^{1\LL \ast } y_u^{ij} \eta^{1\RR}_j .
	\end{equation}  
The same combination of couplings is found in the $ S_1 $ contribution to \gmu; only the $ S_1 $ mass determines the relative size of the contributions. At the best fit point of Fig.~\ref{fig:muon_fit}, the correction to $ y_\mu $ is 50\% of its SM value, and finite naturalness indicates an upper bound of $ M_1 \sim \SI{5}{TeV} $. Finite naturalness of the Higgs mass indicates a similar scale on each of the muoquark masses but is ultimately determined by parameters of the scalar potential that are mostly irrelevant to the low-energy phenomenology.

\section{Conclusion}
NP explanations of the muon anomalies, $ b\to s \mu^+ \mu^- $ and \gmu, are severely constrained by the non-observation of several processes as predicted with the SM.  
New gauged lepton flavor symmetries provide a good organizing principle for LQ explanations, ensuring the recovery of the SM global symmetries at low energies.   
Here we demonstrate the feasibility with a simple model with two muoquarks mediating the anomalies, however, the ideas extend much beyond this one realization. Further directions to take these ideas include the use of the new vector boson as a $ Z' $ mediator for $ b\to s \mu^+ \mu^- $ and the use of alternative symmetries. Recent studies in Ref.~\cite{Greljo:2021npi} explore the space of $ \U(1)_X $ symmetries with the required properties and the possibility of using a light vector boson for \gmu~\cite{Baek:2001kca,Gninenko:2001hx,Altmannshofer:2014pba}.

\acknowledgments 
AET would like to thank the organizers of EPS-HEP2021 for the opportunity to present this work at the
conference. The work of AG and AET has received funding from the Swiss National Science Foundation (SNF) through the Eccellenza Professorial Fellowship ``Flavor Physics at the High Energy Frontier'' project number 186866.
The work of AG is also partially supported by the European Research Council (ERC) under the European Union’s Horizon 2020 research and innovation programme, grant agreement 833280 (FLAY).

\bibliographystyle{JHEP}
\bibliography{GST}

\end{document}